# Discovery of a reversible redox-induced order-disorder transition in a 10-cation compositionally complex ceramic

Dawei Zhang [a], Yan Chen [b], Tianshi Feng [c], Dunji Yu [b], Ke An [b], Renkun Chen [a,c], Jian Luo [a,e,1]

[a] Program of Materials Science and Engineering, University of California San Diego, La Jolla, CA 92093, USA
[b] Neutron Scattering Division, Oak Ridge National Laboratory, Oak Ridge, TN 38731, USA
[c] Department of Mechanical & Aerospace Engineering, University of California San Diego, La Jolla, CA 92093, USA
[e] Department of NanoEngineering, University of California San Diego, La Jolla, CA 92093, USA

**Abstract**

This study discovers a reversible order-disorder transition (ODT) in a 10-cation compositionally complex ceramic, $(Nd_{0.15}Pr_{0.15}Dy_{0.8}Ho_{0.8}Er_{0.8}Ti_{0.2}Yb_{0.1}Hf_{0.1}Zr_{0.1}Nb_{0.8})O_{7-\delta}$, induced via annealing in oxidized *vs.* reduced environments at 1600°C. Notably, the 10-cation oxide remains a homogenous single-phase high-entropy solid solution before and after the ODT in the pyrochlore *vs.* fluorite structure that can be quenched. *In-situ* neutron diffraction reveals the oxygen vacancy formation and atomic displacement during this ODT. The temperature dependence of thermal conductivity is altered by the ODT. This study reveals a new pathway to induce ODT via a redox transition to tailor the properties of compositionally complex fluorite-based oxides.

---

[1] Corresponding Author: jluo@alum.mit.edu

# Graphical Abstract

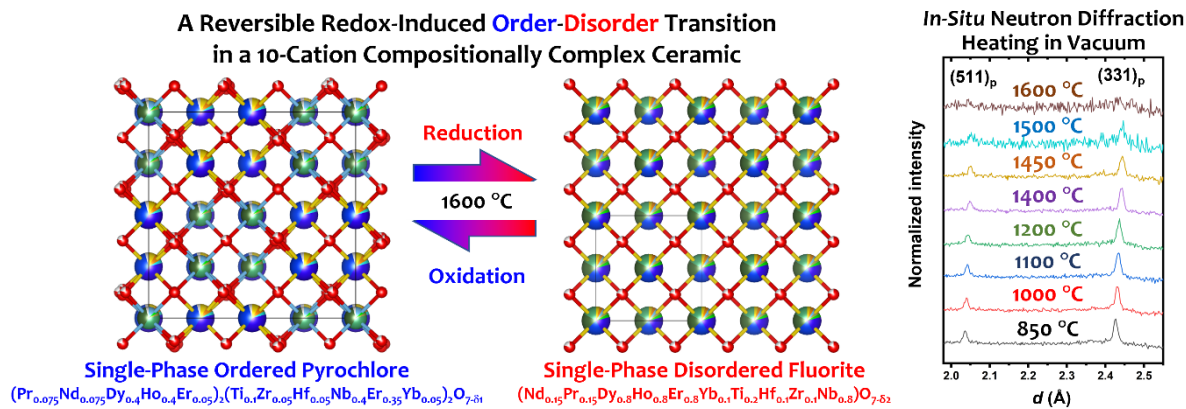

**Highlights:**

- A reversible redox-induced order-disorder transition (ODT) is discovered.

- A 10-cation single-phase ordered pyrochlore oxide is stable at 1600°C in air.

- Reduction at 1600°C induces an ODT to form a single-phase disordered fluorite oxide.

- In-situ neutron diffraction reveals the mechanism of oxygen vacancy induced ODT.

- An ODT can alter the temperature-dependent thermal conductivity curve.

Disorder and order in fluorite-based oxides can influence on a range of physical properties. For example, disorder can increase the ionic conductivities of yttria-stabilized zirconia (YSZ) and $Y_2(Zr_yTi_{1-y})_2O_7$ [1,2], enhance the catalytic activity and stability of Li-O$_2$ batteries [3], and tailor the radiation resistance [4,5]. Long- or short-range pyrochlore and weberite orders can also exist in fluorite-based oxides [6]. The $A_2B_2O_7$ pyrochlore structure can be viewed as a 2 × 2 × 2 superstructure of a base fluorite structure with ordered cations and oxygen vacancies and structural distortion [7]. Prior studies showed that the pyrochlore can transit to a defect fluorite structure, caused by composition [8–12] or temperature [13] induced disorder. In this work, we have further discovered a reversible redox-induced transition between the ordered pyrochlore and disordered fluorite structures.

In 2018, Gild *et al.* reported several single-phase YSZ-like high-entropy fluorite oxides (*e.g.*, $(Zr_{1/5}Hf_{1/5}Ce_{1/5}Y_{1/5}Yb_{1/5})O_{2-\delta}$ and $(Zr_{1/5}Hf_{1/5}Ce_{1/5}Y_{1/5}Gd_{1/5})O_{2-\delta}$) with high cation disorder and reduced thermal conductivities [14]. Since then, high-entropy fluorite oxides [10,14–20] (including rare-earth niobates $RE_3NbO_7$ in the defect fluorite structure [10,17–20]) and fluorite-based pyrochlore oxides [10,15,21–25] have been investigated for potential applications in thermal barrier coatings (TBCs). These high-entropy fluorite and pyrochlore oxides [10,15,17–25] belong to a broader class of high-entropy ceramics (HECs) that have attracted increasing research interests [26], which also include high-entropy rocksalt [27] and pervoskite [28] oxides, borides (of $MB_2$ [29], MB [30], $M_3B_4$ [31], $MB_4$ [32], and $MB_6$ [33,34] metal:boron stoichiometries), silicides [35,36], and carbides [37–39]. To date, most HEC studies are focused on equimolar compositions consisting of typically five (and sometimes four or six) metal cations.

In 2020, it was further proposed to broaden HECs to compositionally complex ceramics (CCCs) to include non-equimolar compositions that can outperform their equimolar counterparts [26,40]. Moreover, compositionally complex fluorite-based oxides (CCFBOs) can also possess points defects (oxygen vacancies and aliovalent cations) and short- and long-range orders, which provide additional dimensions to tailor their thermomechanical properties [10,15,26,40]. Notably, a recent study reported a series of 11-cation single-phase CCFBOs with a composition-induced order-disorder (pyrochlore-fluorite) transition [10]. In a broad perspective, order and disorder can affect (or even control) many thermal and mechanical properties. However, it is not yet demonstrated that an order-disorder transition (ODT) can take place in a CCFBO (or any CCC and HEC in general) of a fixed composition.

In this study, we design a 10-cation oxide $(Nd_{0.15}Pr_{0.15}Dy_{0.8}Ho_{0.8}Er_{0.8}Ti_{0.2}Yb_{0.1}Hf_{0.1}Zr_{0.1}Nb_{0.8})O_{7-\delta}$. This composition is obtained by mixing 20% of a compositionally complex pyrochlore oxide $[(Nd_{0.375}Pr_{0.375}Yb_{0.25})_2(Ti_{0.5}Hf_{0.25}Zr_{0.25})_2O_7]$ with 80% of a high-entropy (fluorite-structured) rare-earth niobate [(DyHoErNb)O$_7$]. The mixture forms a single-phase pyrochlore structure when it is synthesized in air, with the nominal formula: $(Pr_{0.075}Nd_{0.075}Dy_{0.4}Ho_{0.4}\ Er_{0.05})_2(Ti_{0.1}Zr_{0.05}Hf_{0.05}Nb_{0.4}Er_{0.35}Yb_{0.05})_2O_7$

(with smaller cations on the B sites, albeit anti-site defects; see Supplementary Discussion). We denote this 10-cation CCFBO as "10CCFBO$_{0.8Nb}$" for brevity, where the subscript denotes the Nb (or niobate) content. The composition of this 10CCFBO$_{0.8Nb}$ is selected so that it is a barely ordered pyrochlore phase (as it would become a disordered defect fluorite structure in air if we further increased the Nb or fluorite-structured niobate content). This composition design enables us to explore a disordering transition induced by oxygen vacancy generation under a reduced environment, which has not been reported previously, to explore a new route to tailor the order/disorder and properties of CCFBOs (or CCCs in general).

All powders were purchased (from US Research Nanomaterials Inc., Houston, TX), ball milled, and uniaxially pressed into pellets, which were subsequently sintered at 1600 °C for 24 hours in air (following a standard recipe described in a prior report [10]). Specimen densities were measured by the Archimedes method to be ~98% of the theoretical density. We use "pristine" 10CCFBO$_{0.8Nb}$ to refer to (oxidized) as-sintered samples. The pristine 10CCFBO$_{0.8Nb}$ was annealed at 1600 °C for 2 hours in vacuum ($10^{-3}$ mbar) in a graphite furnace (RED DEVIL, RD WEBB, Cambridge, MA) to reduce it to induce an ODT. Subsequently, we re-annealed a reduced 10CCFBO$_{0.8Nb}$ in air at 1600 °C for 12 hours to re-oxidize it to check whether the ODT is reversible. Specimens were characterized by using X-ray diffraction (Miniflex II XRD, Rigaku), a field emission scanning electron microscope (FE SEM, FEI Apero) in conjunction with energy dispersive X-ray spectroscopy (EDS), and a scanning transmission electron microscope (STEM, JEOL 300CF) in conjunction with electron energy loss spectroscopy (EELS). Furthermore, *in-situ* heating neutron diffraction in a vacuum furnace up to 1600 °C was carried out at the VULCAN diffractometer [41], Spallation Neutron Source (SNS) at the Oak Ridge National Laboratory (ORNL). The details of the neutron diffraction experiments and Rietveld refinements can be found in the Supplementary Material. Laser flash analysis (LFA 467 HT HyperFlash, NETZSCH) was used to measure thermal diffusivity and subsequently calculate the thermal conductivity ($k$) using the measured density and estimated heat capacity (following a commonly used procedure described previously [10,15]).

XRD patterns of the pristine, reduced, and re-oxidized specimens showed a reversible redox-induced ODT between ordered pyrochlore and disordered fluorite structures (Figure 1). All three XRD patterns suggested the formation of a single phase (without any detectable secondary phase, as shown in Figure 1) in each case. SEM EDS elemental mapping was used to confirm the compositional homogeneity in all three specimens (Supplementary Fig. S1). The pyrochlore and fluorite phases can be differentiated by the (331) superstructure peak, which were clearly evident for the pristine and re-oxidized specimens, but not detectable in the reduced specimen. Similar to the cases of weakly ordered Gd$_2$Zr$_2$O$_7$ [13,42], the pyrochlore superstructure peaks in the pristine and re-oxidized 10CCFBO$_{0.8Nb}$ were weak because they were barely ordered. We also noticed a reversible color change. As shown in the Supplementary Fig. S2, both the pristine and re-oxidized specimens were brown, but the reduced sample was black (presumably due to the formation of oxygen vacancies). The reduced sample was cross sectioned, grinded, and polished to confirm the color change (reduction) was uniform throughout the whole pellet. We also conducted EELS in STEM to probe the O K edge but only observed a small shift in the reduced specimen (Supplementary Fig. S3) that might be attributed to the reduction (albeit specimen preparation artifacts). In summary, the pristine specimen sintered in air at 1600 °C was ordered pyrochlore, which underwent an ODT to form a disordered defect fluorite phase after reduction at 1600 °C and transformed back to the ordered pyrochlore phase after re-oxidation at 1600 °C.

Prior studies suggested that disordering of $A_2B_2O_7$ pyrochlore can be induced by altering ratio of ionic radii ($r_{A^{3+}}/r_{B^{4+}} < \sim 1.46$ for ternary oxides [8]) or increasing temperature (*e.g.*, in weakly ordered $Gd_2Zr_2O_7$ [13]). In this study, we hypothesize a new redox-induced ODT mechanism where the observed reduction-induced disordering is attributed to the formation of the oxygen vacancies. Here, *ex-situ* benchtop XRD is insufficient to understand this new mechanism. Neutron diffraction can be used to probe structural changes with higher precisions [43–46], including those in high-entropy pyrochlores [47]. Thus, we conducted *in-situ* neutron diffraction experiments at the Vulcan station at the ORNL SNS. Specifically, we heated a pristine (oxidized) specimen in a vacuum furnace with a stepwise heating profile and collected neutron diffraction patterns at 850 °C, 1000 °C, 1100 °C, 1200 °C, 1400 °C, 1450 °C, 1500 °C, and 1600 °C in a sequence. During heating, the gas pressure was kept at $10^{-5}$ mbar to produce a reduced environment. Figure 3 shows the *in-situ* neutron diffraction patterns with an enlarged region showing the pyrochlore (331) and (511) superstructure peaks. With increasing temperature, the superstructure peaks gradually broadened; they completely vanished at 1600°C, indicating a reduction-induced disordering transition (ODT) to form a defect fluorite structure.

Rietveld refinements were conducted to fit the neutron patterns. For the $A_2B_2O_7$ pyrochlore ($Fd\bar{3}m$) structure, we assume that $Yb^{3+}$ and parts of $Er^{3+}$ cations occupy the B site based on the ionic radii (by placing smaller cations on B sites, while maintaining the exact A:B stoichiometry) and the Rietveld refinement fittings (discussed in detail in Supplementary Discussion) so that the pyrochlore chemical formula is $(Pr_{0.075}Nd_{0.075}Dy_{0.4}Ho_{0.4}Er_{0.05})_2(Ti_{0.1}Zr_{0.05}Hf_{0.05}Nb_{0.4}Er_{0.35}Yb_{0.05})_2O_{7-\delta}$. For the disordered fluorite ($Fm\bar{3}m$) phase (for 1600 °C data only), all cations occupy at the same sublattice with the formula $(Nd_{0.15}Pr_{0.15}Dy_{0.8}Ho_{0.8}Er_{0.8}Ti_{0.2}Yb_{0.1}Hf_{0.1}Zr_{0.1}Nb_{0.8})O_{7-\delta}$. With these schemes, the refinement errors were lowered down to $wR_p \approx 0.025$. One fitting example is given in Figure 3(a) and other refinement patterns are documented in Supplementary Fig. S4. In addition to the normal refinement parameters, two major parameters considered here were positional parameter *x* of the O1 (48f) site (at the position *x*, 1/8, 1/8) in the pyrochlore unit cell and the oxygen deficiency δ (that is determined by the occupancies of the three oxygen sites for the pyrochlore structure, as documented and discussed in Supplementary Table S2). As shown in prior reports [1,10], the positional parameter $x$ of the O1 site will move to 0.375 in an ideal pyrochlore to fluorite ODT. The main fitted parameters of refinements are shown in Supplementary Table S2.

Figure 3(b) plots the fitted positional parameter $x$ and oxygen deficiency $\delta$ *vs*. temperature. With increasing temperature, both $x$ and $\delta$ increase. These trends are consistent with prior investigations of the compositionally induced ODT in in $Y_2(Ti_yZr_{1-y})_2O_7$ with increasing $y$ [1,2]. Here, in addition to the displacement of the O1 site (changing of $x$), there are increases in oxygen deficiency. Our *in-situ* neutron experiments suggest that the positional parameter $x$ of the O1 site approaches ~0.367 and oxygen deficiency $\delta$ reaches a significant value of ~0.57, just before the ODT. Here, the average occupancy of the oxygen sites (~0.80; see Supplementary Table S2) is significantly below the nominal value of 0.875 in the defect fluorite structure, creating a large driving force for disordering.

We further measured the thermal conductivities of pristine and reduced $10CCFBO_{0.8Nb}$ (Figure 4). All measurements were conducted from room temperature (25°C) up to 1000°C in an Ar gas environment. Figure 4 shows that both ordered pyrochlore and disordered fluorite $10CCFBO_{0.8Nb}$ exhibit low thermal conductivities and amorphous-like temperature dependence (increasing thermal conductivity with increasing temperature). Interestingly, the reduced $10CCFBO_{0.8Nb}$ has a higher room temperature thermal conductivity (of ~1.1 $W·m^{-1}·K^{-1}$ for the disordered fluorite, *vs*. ~1.05 $W·m^{-1}·K^{-1}$ for the ordered pyrochlore) but a smaller slope with increasing temperature. At 1000 °C, the disordered fluorite $10CCFBO_{0.8Nb}$ has a lower thermal conductivity of ~1.3 $W·m^{-1}·K^{-1}$, compared to ~1.35 $W·m^{-1}·K^{-1}$ for the ordered pyrochlore $10CCFBO_{0.8Nb}$. XRD characterization after the thermal conductivity measurements in Ar up to 1000 °C showed that the measurements did not change the pyrochlore *vs*. fluorite structures of these two specimens (Supplementary Fig. S5). In other words, the order *vs*. disordered structure obtained via redox at 1600 °C can be quenched and preserved during thermal cycling from room temperature to 1000 °C, which offered a unique opportunity to investigate the effects of order *vs*. disorder on thermal conductivity on specimens of an identical composition.

In summary, we have discovered a reversible redox-induced pyrochlore-to-fluorite transition in a 10-cation CCFBO from the ordered pyrochlore to the disordered fluorite structure, while maintaining a single high-entropy solid solution phase before and after this ODT. *In-situ* neutron diffraction revealed that this ODT is induced by the generation of oxygen vacancies. This ODT altered the temperature dependence of the thermal conductivity. This discovery suggests a new pathway to tailor compositionally complex and high-entropy ceramics via redox-induced ODT.

**Acknowledgement:** The work is supported by the National Science Foundation (NSF) in the Ceramics program via Grant No. DMR-2026193. A portion of this research used resources at the Spallation Neutron Source, a DOE Office of Science User Facility operated by the ORNL. The FIB and STEM work utilized the shared facilities at the San Diego Nanotechnology Infrastructure of UCSD, a member of the National Nanotechnology Coordinated Infrastructure (supported by the NSF ECCS-1542148) and the Irvine Materials Research Institute (through UCI CCAM, partially supported by NSF DMR-2011967).

**Figures:**

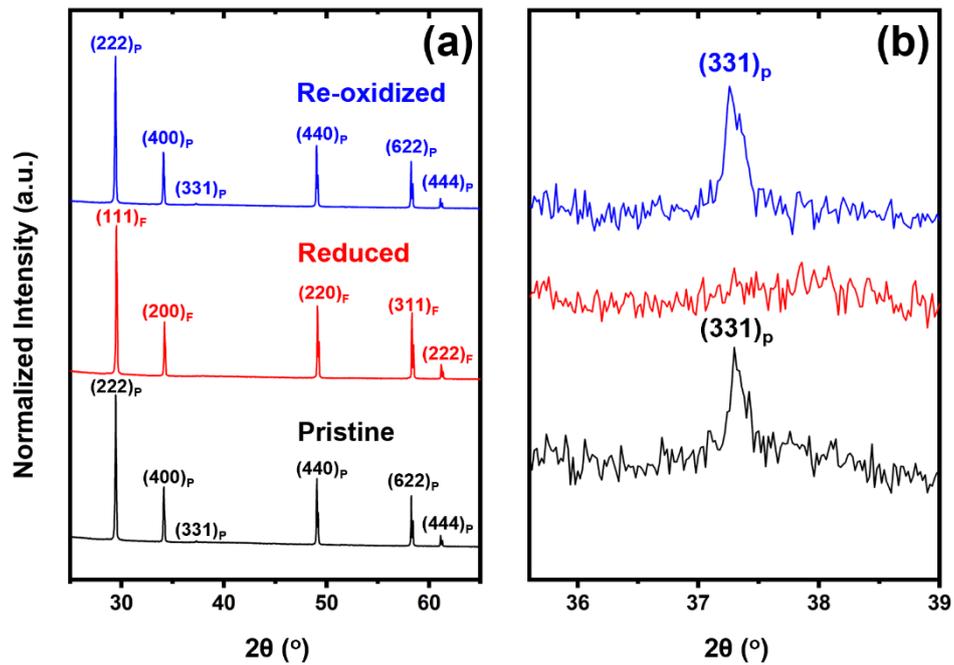

**Figure 1.** XRD patterns of (a) the pristine (synthesized in air), reduced, and re-oxidized specimens, and (b) an enlarged region showing the pyrochlore (331) superstructure peaks. The pristine sample was sintered in air at 1600 °C, which was then annealing at 1600 °C in vacuum (the reduced specimen) and subsequently in air at 1600 °C (the re-oxidized specimen). Here, the pyrochlore $(222)_P$ peak corresponds to the fluorite $(111)_F$ peak, where we use subscript "P" and "F" to denote the phase.

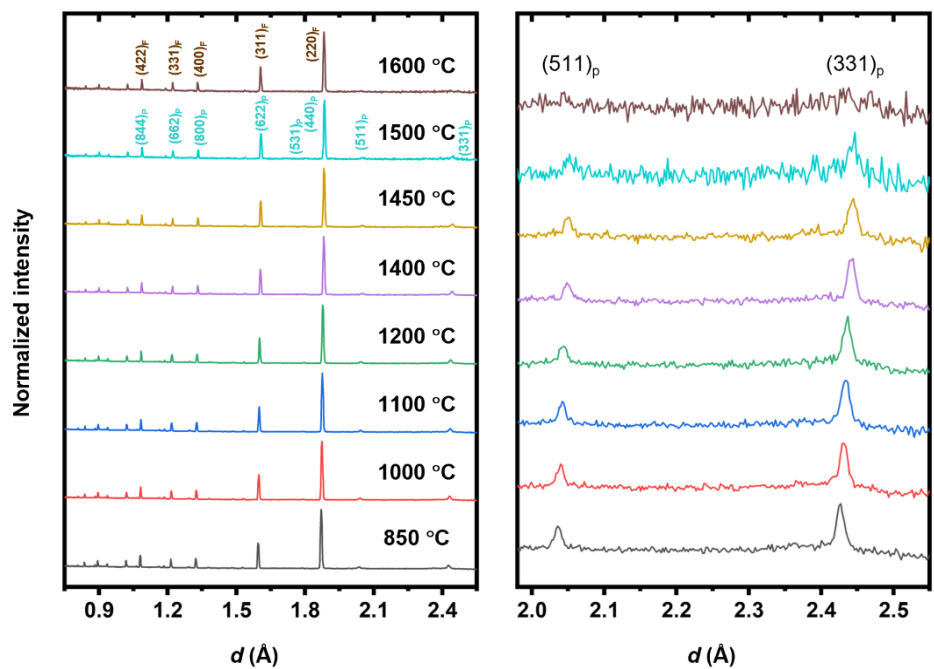

**Figure 2.** In-situ heating (reduction) neutron diffraction patterns collected at different temperatures in vacuum and the enlarged region showing the pyrochlore (331) and (511) superstructure peaks. The peak shift was induced by thermal expansion.

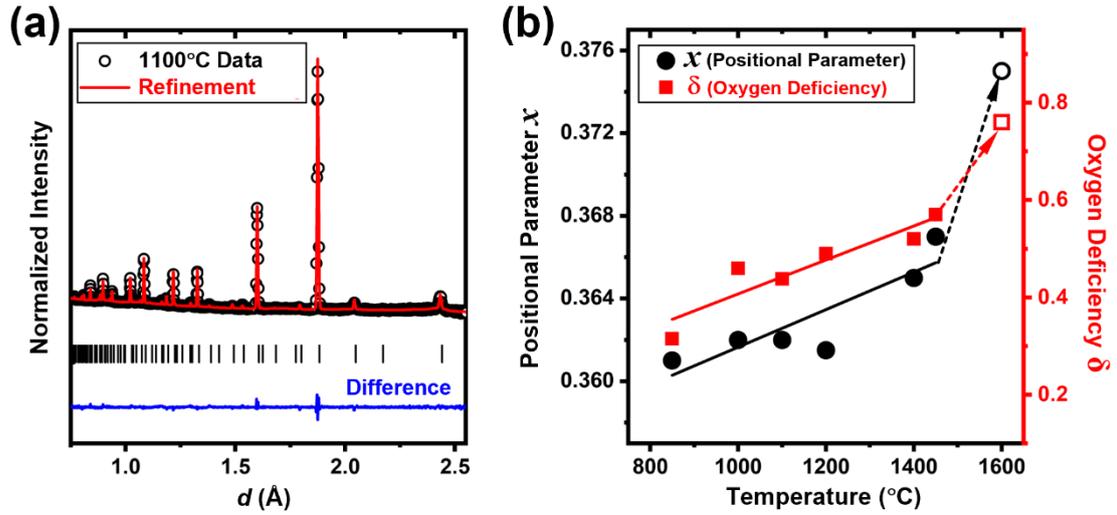

**Figure 3.** (a) An example of the Rietveld refinement of the neutron diffraction pattern (for the measurement at 1100 °C). The fitting error w$R_p$ ≈ 2.5%. Additional cases are shown in the Supplementary Fig. S4. (b) Evolution of positional parameter $x$ and oxygen deficiency δ *vs.* temperature, obtained via Rietveld refinements. The solid symbols were the values refined with the pyrochlore structure and the hollow ones were refined assuming the fluorite structure. The solid lines represent the best fittings for the pyrochlore structures while the dashed lines suggest the transition with the ODT.

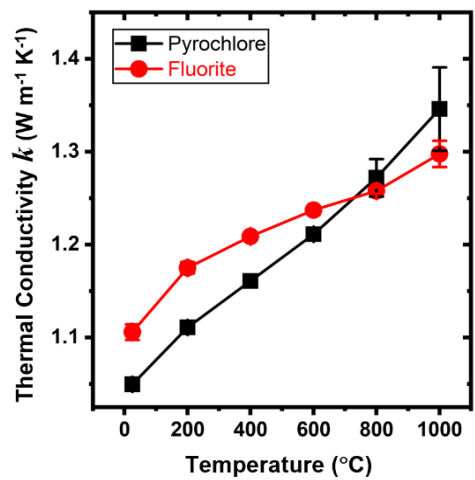

**Figure 4.** Temperature-dependent thermal conductivities of the ordered (pyrochlore) *vs.* disordered (fluorite) specimens of an identical composition. XRD measurements (Supplementary Fig. S5) showed the ordered (pyrochlore) *vs.* disordered (fluorite) structure can be maintained during the measurements up to 1000 °C. The error bars were standard deviations from five measurements at each temperature.